\documentclass[12pt]{article} % or revtex4-2 for physics journal style
\usepackage{amsmath,amssymb}
\usepackage[numbers,sort&compress]{natbib}   
\usepackage{graphicx}
\usepackage{physics}
\usepackage{times}
\usepackage{hyperref}
\usepackage{geometry}
\usepackage[final]{changes}
\title{Planck’s Law from a Classical Free Energy Extremum Involving Fisher Information}
\author{Carlos Gomez-Uribe \\ \small Independent Researcher \\ \small \texttt{cgomez@alum.mit.edu}}
\date{\today}

\begin{document}
\maketitle

\abstract{
We derive Planck’s law from a classical variational principle over probability densities, without invoking quantum states, quantized oscillator energies, or a canonical ensemble over discrete oscillator energy levels. We construct a generalized free energy functional involving entropy and Fisher information, with weights determined by the dimensionless ratio \( \gamma = \hbar\omega / k_B T \). When extremized under a Gaussian ansatz, this functional yields the exact Planck distribution. The only quantum input is a minimal threshold assumption: that an oscillator emits a photon of energy \( \hbar\omega \) only when a thermal fluctuation delivers at least that much energy. We also present a complementary kinetic derivation, based on threshold-activated thermal \emph{\added{emission} cascades}, that yields the same result through classical stochastic reasoning. Together, these approaches suggest that Planck’s law — long considered a hallmark of quantum theory — may instead arise from classical thermodynamic principles supplemented by minimal constraints. This reframing has potential implications for understanding the emergence of quantum behavior from classical statistical systems.
}

\section{Introduction}
Black-body radiation refers to the electromagnetic emission from an object in thermal equilibrium. A black body is an idealized system that absorbs all incident radiation and re-emits energy determined solely by its temperature. This emission results from thermal motion of charged particles that stochastically excite electromagnetic modes. At equilibrium, absorption and emission balance, producing a universal spectral distribution. Black-body spectra have been confirmed across a wide range of systems and scales. The cosmic microwave background (CMB), measured by FIRAS on the COBE satellite, matches a Planck spectrum at \( T = 2.725\,\text{K} \) to within 50 parts per million~\cite{Fixsen1996}. Stellar radiation, such as the Sun’s, is well-approximated by a black body at 5777 K. In the lab, small reflective cavities with apertures generate nearly ideal spectra at controlled temperatures. These measurements tightly constrain any theoretical derivation.

The failure of classical electrodynamics to explain black-body radiation—the so-called ultraviolet catastrophe—revealed a deep inconsistency between thermodynamics and Maxwell’s theory. Planck resolved this by postulating discrete \(\hbar \omega\) energy exchanges, a step he initially viewed as a mathematical convenience. This idea marked the inception of quantum theory and catalyzed a broader shift toward quantized models in atomic and radiation physics, including Bohr’s model of the hydrogen atom and Einstein’s photoelectric effect. Planck’s law, a foundational result in quantum theory, expresses the spectral energy density \( \rho(\omega, T) \), the energy per unit volume per unit angular frequency, as the product of two quantities:
\begin{equation}
    \rho(\omega, T) = \langle E(\omega) \rangle \cdot n_\omega, \label{eq:PlanckSpectra}
\end{equation}
where
\begin{equation}
    \langle E(\omega) \rangle = \frac{\hbar \omega}{e^{\gamma} - 1}, \quad 
    n_\omega = \frac{\omega^2}{\pi^2 c^3}, \quad 
    \gamma = \frac{\hbar \omega}{k_B T}.\label{eq:PlanckSpectra2}
\end{equation}
Here \( \langle E(\omega) \rangle \) is the average energy per mode of frequency \( \omega \), and \( n_\omega \) is the density of electromagnetic modes per unit volume per unit angular frequency, and $\gamma$ is the ratio of the two energy scales in the system: photon emission and thermal fluctuations. The dimensionless energy ratio \(\gamma\) governs the crossover between classical and quantum regimes and will play a central role in our formulation. Note that the expression for \( \langle E(\omega) \rangle \) excludes the zero-point energy \( \frac{1}{2} \hbar \omega\), as black-body radiation depends only on energy differences. Integrating \( \rho(\omega, T) \) over all frequencies yields the total energy density, known as the Stefan–Boltzmann law:
\(    u(T) =  \frac{\pi^2 k_B^4}{15 \hbar^3 c^3} T^4.\)

Planck modeled the electromagnetic field inside a cavity as a collection of harmonic oscillators, each corresponding to a mode of vibration with frequency \(\omega\). This idealization reflects the linearity of Maxwell's equations and the standing-wave nature of cavity modes, leading—under quantization—to discrete energy levels \( E_n = (n + 1/2)\hbar \omega \) obtained by solving the Schrödinger equation. A thermal ensemble over these states yields the average mode energy. Alternative approaches, such as Einstein’s theory of spontaneous and stimulated emission, still assume quantized states and transitions~\cite{Einstein1917}. In contrast, here we derive Planck’s law from a variational principle over probability densities in position space. We define a generalized free energy functional incorporating expected potential energy, Shannon entropy, and Fisher information~\cite{CoverThomas}. The weights of the entropy and Fisher terms depend on the dimensionless quantum-to-thermal ratio \( \gamma \), and encode the competition between thermal fluctuations and quantum localization. By extremizing this free energy using a Gaussian ansatz for the position distribution of a harmonic oscillator, we recover the full Planck expression for \( \langle E(\omega) \rangle \). A complementary derivation has been proposed by Cetto and de la Peña \cite{cetto2017real}, who recover Planck’s law from a thermodynamic analysis of a harmonic oscillator subjected to superimposed zero-point and thermal fluctuations. Their approach reproduces the correct average energy per mode by statistically averaging over phase and amplitude variables of the combined motion. While their model is rooted in classical stochastic dynamics involving background fields, our derivation follows from a variational principle over configuration-space distributions alone.

Our framework does not rely on quantum states, quantized oscillator levels, or Bose–Einstein statistics. The only explicitly quantum ingredient is the threshold assumption that photon emission requires energy \(\hbar \omega\). \added{The present framework does not attempt to derive the existence of photons or the relation $E_\gamma=\hbar\omega$ from purely classical dynamics. Instead, we treat $E_\gamma=\hbar\omega$ as an experimentally established constraint on radiative emission and ask a narrower question: \emph{given this minimal threshold, can the full Planck factor $(e^\gamma-1)^{-1}$ emerge from a classical free-energy extremum over continuous densities without quantized oscillator levels or Bose counting?}} \replaced{Our}{The} inclusion of Fisher information is motivated by information geometry and steady-state reasoning, and while it shares mathematical structure with quantum variational principles, it is employed here purely in a classical setting. This suggests an alternative physical foundation for black-body radiation, rooted in information geometry and thermodynamic variational principles.\footnote{\added{We use \emph{information geometry} to mean the standard view of probability densities as a statistical manifold whose local metric is induced by  the Fisher information, which measures localization/roughness of probability densities in configuration space.  We refer to the Fisher term in Eq.~\eqref{eq:general_free_energy} as a \emph{geometric constraint} because it penalizes the \emph{shape} of the steady-state density rather than representing a thermal driving contribution.}} This direction complements earlier efforts to recover Planck’s law from classical frameworks, including models based on stochastic electrodynamics~\cite{Boyer1969,DeLaPena2015} and statistical reinterpretations of Maxwell theory~\cite{Franca1996}.

\section{Free Energy Extremization Principles}

The equilibrium distribution of a classical system in contact with a thermal bath minimizes the Helmholtz free energy functional,
\begin{equation*}
    F_{\text{classical}}[p] = \langle U \rangle - k_B T \cdot H(p),
\end{equation*}
where \( p(x) \) is a probability density over configuration space, \( \langle U \rangle = \int p(x) U(x) dx \) is the expected potential energy, and \( H(p) = -\int p(x) \log p(x) dx \) is the Shannon entropy. In contrast, the states of a quantum system extremize the free energy
\begin{equation}
    F_{\text{quantum}}[p] = \langle U \rangle + \frac{\hbar^2}{8m} J(p), \label{eq:FQM}
\end{equation}
where \( J(p) = \int \frac{(\nabla p(x))^2}{p(x)} dx \) is the Fisher information associated with the position distribution. For a broad class of systems, extremizing this functional recovers the time-independent Schrödinger equation.  See Appendix B for a derivation of this expression from the quantum Hamiltonian and its interpretation as a decomposition of the kinetic energy into Fisher information and phase flow — e.g., as in the Madelung formulation.

This motivates a variational framework whose solutions interpolate between classical and quantum limits in the harmonic oscillator case. We propose the generalized free energy:
\begin{equation}
    F_\gamma[p] = \langle U \rangle - k_B T \cdot f(\gamma) \cdot H(p) + g(\gamma) \cdot \frac{\hbar^2}{8m} J(p),
    \label{eq:general_free_energy}
\end{equation}
where \( f(\gamma) \), \( g(\gamma) \) are scalar functions that weight entropy and information contributions respectively. The structure of Eq.~\ref{eq:general_free_energy} is not derived from microscopic dynamics but rather proposed as a thermodynamic interpolation between classical and quantum regimes. This framework seeks a steady-state configuration rather than time-evolving dynamics. We interpret the extremum as representing a thermodynamically stable energy distribution. In the next section, we motivate the specific forms of \( f(\gamma) \) and \( g(\gamma) \) by analyzing thermal excitation thresholds, and then solve the variational problem using a Gaussian ansatz, since it is known to extremize both entropy and Fisher for quadratic potentials.

\subsection{Entropy Scaling with Energy Quanta}
We begin with the entropy term. In classical statistical mechanics, entropy is weighted by the thermal energy scale \( k_B T \). In our setting, however, thermal fluctuations are not sufficient to induce radiation unless they exceed a finite energy threshold. Specifically, the oscillator emits a photon only when it absorbs energy \( E \geq \hbar \omega \), since photon emission at frequency \( \omega \) requires a minimum quantum of energy \( \hbar \omega \). Thermal kicks with lower energy do not result in emission and instead contribute only to background fluctuations.

We model thermal energy kicks as continuous energy transfers drawn from a Boltzmann distribution. In classical statistical mechanics, the probability that a system in thermal equilibrium at temperature \( T \) possesses energy \( E \) is proportional to \( e^{-E / k_B T} \). This holds for systems exchanging energy with a large reservoir, and corresponds to the energy distribution in the canonical ensemble. Accordingly, we assume that the probability density for individual thermal kicks to deliver energy \( E \) is exponential with mean \( k_B T \):
\begin{equation*}
    P(E) = \frac{1}{k_B T} e^{-E / k_B T}.
\end{equation*}
Under this model, the probability that a single thermal fluctuation delivers enough energy to produce photon emission at frequency \( \omega \) — i.e., energy exceeding \( \hbar \omega \) — is:
\begin{equation*}
p_{\mathrm{th}} = \mathbb{P}(E > \hbar \omega) = e^{-\gamma}.
\end{equation*}

Since only a fraction \(p_{\mathrm{th}} \) of kicks are effective in producing emission, we argue that the entropy term in the free energy should be weighted by the relevant energy scale for emission events: \( \hbar \omega \). Thus we set
    \(f(\gamma) = \gamma \) since that makes the entropy term in the free energy \( \hbar \omega H(p)\) rather than the typical \(k_B T H(p).\)
This ensures that entropy cost reflects the size of the required excitation, and is independent of the thermal environment's effective temperature.

\subsection{Fisher Information Scaling and Failed Kicks}

Next, we turn to the Fisher information term. Fisher information penalizes sharp gradients in the probability distribution, and can be viewed as a measure of localization. We assume that failed thermal kicks—those below threshold—perturb the system without causing photon emission. If each failed kick leads to a small random perturbation in position, then the net sharpness of the probability distribution \( p(x) \) increases with the number of such kicks.

Given that successful excitation occurs with probability \( p_{\mathrm{th}} = e^{-\gamma} \), the expected number of failed kicks per successful one is:
\begin{equation*}
    n_T = \frac{1}{p_{\mathrm{th}} } - 1 = e^{\gamma} - 1.
\end{equation*}
We propose that the number of failed kicks \( n_T \) is proportional to the scale of the perturbation to \( \nabla p(x) \), and hence \( J(p) \propto n_T^2 \).
To preserve the interpretation of the Fisher term as a geometric constraint rather than a thermodynamic driver, we require that it not scale with temperature in the high-temperature limit. To avoid an unphysical growth of the Fisher contribution in the high-temperature limit, we take the Fisher
prefactor to scale inversely with the square of the expected number of sub-threshold (failed) attempts per
successful threshold crossing. Since $n_T = e^\gamma-1$, this motivates a scaling
$1+g(\gamma) \propto (e^\gamma-1)^{-2}$.
We choose the constant offset so that $g(\gamma)\ge -1$ and the extremal solution approaches the $T\to 0$
ground-state variance $\sigma^2 \to \hbar/(2m\omega)$, while still reproducing the correct thermal scaling for
$\gamma\ll 1$.

The final form of \(g(\gamma)\) that satisfies both the zero-point limit and reduces to the classical thermal result at small \( \gamma \) is:
\begin{equation}
    g(\gamma) = \left(\frac{2}{e^{\gamma} - 1}\right)^2 - 1. \label{eq:g}
\end{equation}
The prefactor of 2 in the Fisher term ensures that in the high-temperature limit \( (\gamma \ll 1) \), the variational free energy is minimized by a distribution whose variance reproduces the classical equipartition result \( \langle U \rangle = \frac{1}{2} k_B T \), up to the $T$-independent zero-point offset. Interestingly, even though our functional does not reduce to the quantum variational principle when \( T \to 0 \) at fixed \( \omega \), the minimizing distribution still converges to the quantum mechanical ground state in this limit. This can be seen directly from Eqs.~\ref{eq:optimal_var} and \ref{eq:mean_potential} below, which recover the quantum variance and zero-point energy as \( T \to 0 \), despite the presence of the entropy term in our free energy. Thus, while the functionals differ, their minima coincide in this limit. The function \( g(\gamma) \) changes sign at \( \gamma = \log 3 \), corresponding to a crossover temperature \( T_c = \hbar \omega / (k_B \log 3) \). For \( \gamma < \log 3 \), the Fisher term is positive and penalizes localization, consistent with thermal smoothing; for \( \gamma > \log 3 \), it becomes negative and favors sharper localization, corresponding to reduced thermal fluctuations. This structure enables the Fisher term to encode the transition from thermal to quantum-dominated regimes. This assumes the oscillator samples energy fluctuations frequently enough to reach steady state, with strongest coupling near classical turning points. While no explicit dynamics are modeled, the steady-state variational principle may be viewed as characterizing the long-time average distribution resulting from thermal interactions.

With these choices, our free energy becomes
\begin{equation}
    F[p] = \langle U \rangle - \hbar \omega  H(p) + \biggr[\left(\frac{2}{e^{\gamma} - 1}\right)^2 - 1 \biggr]  \frac{\hbar^2}{8m} J(p).
    \label{eq:general_free_energy_SHO1}
\end{equation}
 We restrict our variational search to normalized Gaussians with zero mean, which are known to extremize both the entropy and Fisher information for quadratic potentials. Our goal is not to classify all extrema, but to show that a natural analytic one yields Planck’s law. Full details of the extremization appear in Appendix C, including confirmation that the Gaussian ansatz yields an extremal solution of Eq.~\ref{eq:general_free_energy_SHO1}.

The SHO potential \( U(x) = \frac{1}{2} m \omega^2 x^2 \) yields the expected potential energy \(  \langle U \rangle = \frac{1}{2} m \omega^2 \sigma^2 
\) under our Gaussian assumption. The Shannon entropy and Fisher information of the Gaussian distribution are respectively
\( H(p) = \frac{1}{2} \log(2\pi e \sigma^2), \)  and  \(  J(p) = \frac{1}{\sigma^2}.\) Substituting into Eq.~\ref{eq:general_free_energy_SHO1}, the free energy becomes a function of \( \sigma^2 \):
\begin{equation}
    F(\sigma^2) = \frac{1}{2} m \omega^2 \sigma^2 - \frac{1}{2}\hbar \omega  \log(2\pi e \sigma^2) +  g(\gamma)\frac{\hbar^2}{8m}  \frac{1}{\sigma^2}.
    \label{eq:general_free_energy_SHO}
\end{equation}
To find the optimal variance, we extremize this expression with respect to \( \sigma^2 \). Taking the derivative with respect to  \( \sigma^2 \), setting it to zero, and solving the resulting quadratic equation yields:
\begin{equation}
    \sigma^2 = \frac{1}{2}\frac{\hbar}{m \omega} \left( 1 \pm \sqrt{ 1 + g(\gamma) } \right) = \frac{1}{2}\frac{\hbar}{m \omega} \left(1 + \frac{2}{e^\gamma - 1} \right),\label{eq:optimal_var}
\end{equation}
where we take the positive branch to ensure that $\sigma^2 > 0,$ as well as the appropriate high temperature limit. Substituting into the expected potential energy yields
\begin{equation}
    \langle U \rangle = \frac{1}{2}\biggr(\frac{\hbar \omega}{2} + \frac{\hbar \omega}{ e^\gamma - 1} \biggr).\label{eq:mean_potential}
\end{equation}
In the quantum ground state, the average kinetic and potential energies are equal, so \( \langle E \rangle = 2 \langle U \rangle = \hbar \omega / 2 \), as expected.

We now address how to interpret the total energy per oscillator, given that our variational principle is defined only over the probability distribution \( p(x) \). In classical thermodynamics, the total energy of a harmonic oscillator is the sum of kinetic and potential energies, and these are equal in equilibrium due to equipartition. Since our free energy is not canonical, equipartition cannot be assumed. We instead extend the model to phase space, assuming a factorized distribution \(p(x, v) = p(x) p(v)\), where \(v\) is velocity. This assumption is justified in classical systems, such as the steady state of the Langevin equation for a harmonic oscillator, where phase-space distributions factorize in equilibrium as described in standard treatments of stochastic thermodynamics~\cite{Kadanoff2000}. While such a factorization is not generally valid in quantum mechanics due to the Fourier relationship between position and momentum, our model does not invoke quantum states or wavefunctions. We therefore adopt this classical assumption as a thermodynamic idealization. The phase-space structure is thus used as a computational scaffold to recover total energy, not as a literal dynamical model.

Under this assumption, the total free energy functional becomes:
\begin{align*}
F[p(x,v)] &= \langle U(x) \rangle + \langle T(v) \rangle  - \hbar \omega \cdot \left( H[p(x)] + H[p(v)] \right) \nonumber \\
&\quad + \frac{\hbar^2}{8m} \cdot \left( g(\gamma) J[p(x)] + g(\gamma) \omega^2 J[p(v)] \right).
\end{align*}
Here, the potential in velocity space is \( \frac{1}{2} m v^2 \), and the Fisher information term \( J[p(v)] \) carries an additional factor of \( \omega^2 \) to preserve dimensional consistency, since Fisher information scales inversely with the square of the coordinate. Given this factorized structure and symmetry, the variational problems for \( p(x) \) and \( p(v) \) are structurally identical. Therefore, the optimal distributions have equal shape, and the resulting variances satisfy \( \langle T \rangle = \langle U \rangle \). This equipartition of energy emerges naturally from the internal symmetry of the free energy, not from an external thermodynamic postulate. We thus identify the total energy per oscillator as:
\begin{equation}
\langle E \rangle = \langle U \rangle + \langle T \rangle = 2 \langle U \rangle = \frac{\hbar \omega}{2} + \frac{\hbar \omega}{e^\gamma - 1}.\label{eq:mean_E}
\end{equation}
Note that this energy is computed directly from expectation values over the extremal distribution, and is not derived from the variational eigenvalue \(\lambda\); see Appendix C.
Only the second term in \(\langle E \rangle \) contributes to black-body radiation, since energy differences drive emission, and the \( \hbar \omega / 2 \) zero-point term remains fixed as \( T \to 0 \). The radiative energy per mode is therefore:
\begin{equation*}
\langle E_{\text{rad}} \rangle = \langle E \rangle - \lim_{T \to 0} \langle E \rangle = \frac{\hbar \omega}{e^\gamma - 1},
\end{equation*}
which multiplied by \(n_\omega\) recovers the Planck blackbody spectra in Eqs.~\ref{eq:PlanckSpectra} and \ref{eq:PlanckSpectra2}.

\section*{Discussion}

We have shown that Planck’s radiation law can be derived from a generalized free energy functional over classical configuration-space distributions, without invoking oscillator quantized states, energies, or thermal ensembles.  The only quantum ingredient is a threshold rule: radiation occurs only when thermal fluctuations supply at least \( \hbar \omega \) of energy, consistent with photon emission in discrete quanta. While previous work has connected Fisher information to quantum behavior~\cite{Frieden2004}, our contribution lies in identifying physically motivated scaling functions \( f(\gamma) \) and \( g(\gamma) \) that interpolate between classical and quantum regimes. The resulting extremum yields the exact Planck spectrum without invoking quantized states. The Fisher term in our free energy shares mathematical structure with quantum free energy, but here plays a classical, information-theoretic role.

Our derivation contributes to a growing body of work exploring how classical principles might recover blackbody radiation spectra traditionally attributed to quantum mechanics. Notably, Boyer~\cite{Boyer1969,Boyer2010} and de~la~Peña, Cetto, and Valdés-Hernández~\cite{DeLaPena2015} have shown that Planck’s spectrum can be recovered from classical electrodynamics when augmented with a Lorentz-invariant zero-point field, a central idea in stochastic electrodynamics (SED). Boyer derives the spectrum by combining relativistic invariance with classical zero-point radiation. A comprehensive formulation of SED by de~la~Peña and collaborators demonstrates how quantum phenomena—including Planck’s law—can emerge from classical dynamics coupled to stochastic zero-point fields. Cetto and de~la~Peña~\cite{cetto2017real} show more recently that thermodynamic analysis of an oscillator with superimposed zero-point and thermal oscillations leads to Planck's law. França et al.~\cite{Franca1996} independently recover the Planck spectrum from classical statistical mechanics and Maxwell’s equations, showing that Bose–Einstein-like energy distributions can arise without invoking quantized states. These works treat the particle as embedded in a stochastic field, while our approach considers position distributions without explicit reference to an underlying medium. Both frameworks, however, suggest that Planck’s spectrum can arise from classical dynamics when appropriately constrained. Boyer later also emphasizes the historical and conceptual significance of these classical routes, which remain largely absent from modern treatments~\cite{Boyer2018}. 

In contrast, our approach is grounded not in field dynamics or stochastic processes, but in thermodynamic variational principles and information geometry. We construct a generalized free energy functional incorporating entropy and Fisher information, with coefficients that depend on the dimensionless ratio \( \gamma = \hbar\omega / k_B T \). These coefficients are motivated by threshold activation arguments and encode the interplay between thermal fluctuations and quantum-scale emission thresholds. Our formulation operates directly in configuration space and makes no assumptions about quantized energy levels, photon statistics, or background fields. The Planck spectrum emerges as the extremum of this classical free energy, under a simple Gaussian ansatz. In contrast to quantum variational principles, where the zero-point energy arises solely from a Fisher-like localization cost (cf. Eq.~\ref{eq:FQM}, the Fisher-based quantum free energy), our framework includes a persistent entropy term scaled by \(\hbar\omega\). At zero temperature, the balance between this entropy and the Fisher term — whose coefficient becomes negative — yields a nonzero steady-state energy of \(\hbar\omega/2\). This structure implicitly encodes zero-point effects without invoking an explicit vacuum field, replacing it with information-theoretic constraints in configuration space.

While Fisher information has been previously proposed as a bridge to quantum mechanics—notably in Frieden’s work~\cite{Frieden2004}—our approach is conceptually distinct. We do not invoke measurement-based inference, information-theoretic axioms, or statistical estimation principles. Instead, Fisher information enters as a geometric regularizer in a classical thermodynamic variational framework. This avoids speculative postulates and remains grounded in physical assumptions: threshold activation and spatial localization under thermal fluctuations. Our results show that Fisher information can play a role in emergent quantum behavior without adopting Frieden’s broader interpretive apparatus.

In our formulation, the classical limit corresponds to \( \gamma \ll 1 \), where thermal fluctuations dominate over the emission threshold. In this regime, the generalized free energy in Eq.~\ref{eq:general_free_energy_SHO} differs structurally from the classical entropy-based functional, yet both yield the same minimizing variance \( \sigma^2 = k_B T / (m \omega^2) \). This ensures consistency with classical Boltzmann statistics even though our functional retains explicit \( \hbar \omega \)-scaled entropy and a nonvanishing Fisher term. The classical behavior thus emerges not from eliminating quantum terms, but from the limiting behavior of their coefficients.

\begin{table*}[h]
\centering
\begin{tabular}{|p{2cm}|p{3.7cm}|p{3.7cm}|p{3.7cm}|}
\hline
\textbf{Approach} & \textbf{Key Assumptions} & \textbf{Main Tools} & \textbf{Quantum Element} \\
\hline
\textbf{Planck (1901)} & Quantized energy in units of \( \hbar \omega \); oscillator ensembles & Boltzmann entropy \( S = k_B \log W \), combinatorics & Postulates energy quantization to match data \\
\hline
\textbf{Einstein (1917)} & Two-level atoms with energy gap \( \hbar \omega \) & Detailed balance, rates \( A, B \), classical field & Discrete atomic states; field remains classical \\
\hline
\textbf{Bose (1924)} & Indistinguishable photons; no atom model needed & Counting in phase space; Bose statistics & Full quantum photon statistics \\
\hline
\textbf{Jaynes (1957)} & Maximum entropy over discrete energy levels & Entropy maximization; inference logic & Assumes \( E_n = n \hbar \omega \) energy levels \\
\hline
\textbf{Boyer (2010)} & Classical zero-point radiation; relativistic invariance & Thermal equilibrium in accelerating frames & Zero-point field with scale set by \( \hbar \) \\
\hline
\textbf{Cetto \& de la Peña (2017)} & Classical oscillator driven by zero-point + thermal fields & Statistical averaging over stochastic oscillations; thermodynamic arguments & Zero-point energy added as background fluctuation \\
\hline
\textbf{França et al. (1996)} & Maxwell theory + classical thermal field & Stochastic classical energy exchange; Bose-like stats & No discrete levels; uses continuous classical field \\
\hline
\textbf{This Work (Variational)} & Classical oscillator; emission only if \( E > \hbar \omega \) & Extremizes free energy with entropy and Fisher info & Threshold energy for emission only \\
\hline
\textbf{This Work (Kinetic)} & Thermal kicks with \( E > \hbar \omega \) trigger \added{emission} cascades & Classical stochastic kinetics, detailed balance & Threshold energy for emission only \\
\hline
\end{tabular}
\caption{Classical and quantum derivations of Planck’s law. Our two formulations recover the spectrum from classical principles.}
\label{tab:derivations}
\end{table*}

All standard quantum derivations of Planck’s law assume discrete energy levels or quantized radiation. Planck’s original model~\cite{Planck1901} introduced oscillator quantization and ensemble averaging. Einstein’s approach~\cite{Einstein1917} invoked transition probabilities between discrete atomic levels. Bose’s derivation~\cite{Bose1924} relied on indistinguishable photons and Bose–Einstein statistics. Jaynes~\cite{Jaynes1957} emphasized entropy maximization over discrete mode energies. A formally similar result also emerges by minimizing the free energy \( \langle E \rangle - k_B T H(p_n) \) over discrete energy distributions \( p_n \propto e^{-n \hbar \omega / k_B T} \). These ensemble-based derivations all assume quantized energy spectra. In contrast, our formulation is entirely continuous and classical, requiring only that photon emission occurs when thermal fluctuations deliver at least \( \hbar \omega \) of energy. Absorption, by contrast, is unrestricted: thermal kicks can incrementally increase a mode’s energy without threshold. This asymmetry still satisfies detailed balance and leads to the Planck spectrum through equilibrium dynamics.

Finally, our kinetic interpretation—developed in Appendix A—provides an alternative classical route to the Planck's law, based on thermal \replaced{emission}{activation} cascades. \added{
An \emph{emission cascade} is a burst process in which an above-threshold activation event opens a short-lived ``emissive'' window. During this window, emission attempts occur repeatedly and each attempt succeeds only if the instantaneous thermal energy transfer exceeds $\hbar\omega$; the window closes at the first sub-threshold attempt. This produces a geometrically distributed burst size (Appendix D provides a minimal stochastic realization and simulation).} This framework suggests potential experimental signatures in low-temperature, single-mode thermal radiation systems. If energy enters electromagnetic modes in discrete bursts triggered by supra-threshold fluctuations, then photon emission may exhibit bunched statistics or non-Poissonian correlations. Future time-resolved measurements could test these predictions and clarify the microscopic dynamics underlying thermal radiation.  A summary of major derivations, including our variational and kinetic formulations, is provided in Table~\ref{tab:derivations}. Our results suggest that Planck’s law may be viewed not as a direct consequence of quantization, but as the natural equilibrium outcome of classical systems governed by threshold constraints and variational structure. In particular, the cascade mechanism introduces the possibility of experimental signatures—such as photon clustering or correlated emission statistics—that could provide empirical tests of this classical activation-based model of radiation. More broadly, these results hint at a deeper connection between thermodynamic structure, information geometry, and quantum thermodynamics.

\section*{Acknowledgments}

We thank Ana María Cetto and Luis de la Peña for their insightful comments on earlier drafts of this work.  CGU is also grateful to John W. Bush for sparking his interest in the foundations of quantum mechanics through multiple conversations around the fascinating walking-droplet experiments. This research was conducted independently and received no external funding.

\bibliographystyle{apsrev4-2}
\bibliography{refs}

\appendix

\section*{Appendix A: Classical Kinetic Derivation of Planck’s Law via Emission Cascades}
In this appendix, we present a classical kinetic model of radiation that offers an alternative derivation of Planck’s law. This approach is physically motivated by threshold-triggered \replaced{emission}{energy} cascades initiated by thermal fluctuations.
We consider energy exchange between a thermal bath and a single electromagnetic (EM) mode of frequency \(\omega\). The key physical mechanism is a stream of thermal interaction events, or ``kicks," delivered by atoms in the surrounding body. These kicks occur at an average rate \(r\) that may depend on \(T\), and their energies are drawn from a Boltzmann distribution with mean \(k_B T\). Each kick represents a discrete opportunity for energy transfer, either into or out of the EM mode. 
\added{The kinetic balance model here describes the \emph{thermally activated, radiative} energy stored in the mode,
i.e.\ the temperature-dependent part above the $T\to 0$ baseline.}

To inject energy into the \added{EM} mode, a kick must deliver at least \(\hbar\omega\). The probability of this occurring is \(p_{\mathrm{th}}  = e^{-\gamma}\), where \(\gamma = \hbar\omega / k_B T\). A kick that exceeds this threshold initiates an \emph{emission cascade}: a sequence of emissions continues as long as subsequent kicks also exceed the threshold, and ends with the first sub-threshold kick. Equivalently, we idealize an above-threshold kick as opening a short-lived emissive window during which multiple emission attempts occur; each attempt independently exceeds threshold with probability $e^{-\gamma}$, and the window closes on the first sub-threshold attempt. This process produces a geometrically distributed number of emissions per cascade with mean \(\langle N \rangle = 1 / (1 - p_{\mathrm{th}} )\), and average energy input per cascade \(\langle E_{\text{cascade}} \rangle = \hbar\omega / (1 - e^{-\gamma})\). Only a fraction \(e^{-\gamma}\) of all kicks initiate cascades, so the average cascade rate is \(\alpha = r e^{-\gamma}\). The total energy inflow rate is then \(\alpha \cdot \langle E_{\text{cascade}} \rangle = r e^{-\gamma} \cdot \hbar\omega / (1 - e^{-\gamma})\).

Energy also leaves the EM mode when atoms absorb energy from it. We assume that absorption occurs during the same type of thermal interaction events — i.e., atoms are only “open” to absorbing energy during kicks — but unlike emission, there is no energy threshold for absorption. Any kick can remove energy from the mode, with probability and magnitude governed by detailed balance. Since these kicks occur with rate \(r\), and absorption is not threshold-limited, the rate of energy loss is proportional to the stored energy: \(r \cdot \langle E \rangle\). At steady state, energy input and output must balance. This yields:
\[\langle E \rangle = e^{-\gamma} \cdot \frac{\hbar\omega}{1 - e^{-\gamma}} = \frac{\hbar\omega}{e^{\gamma} - 1},\]
which is precisely Planck's expression for the average energy per mode at temperature \(T\). In this model, the EM mode serves as an intermediary between thermal energy from the atomic bath and radiated photons. 
While we compute the steady-state mode energy by balancing thermal input and output rates, actual photon emission requires that a fraction $\eta$ of this energy escapes the system. 
In the idealized blackbody limit ($\eta = 1$), all energy injected into the EM mode eventually radiates, yielding Planck’s law. 
For real systems with partial emissivity, $\eta < 1$ leads to reduced radiative output but does not alter the internal cascade dynamics.

A crucial aspect of this derivation is the presence of \emph{\added{emission} cascades}.\replaced{Without cascades, suppose each kick injects at most one quantum: with probability $p_{\mathrm{th}}=e^{-\gamma}$
a kick adds energy $\hbar\omega$, otherwise it adds zero.  Then the mean input power is
$r\,p_{\mathrm{th}}\,\hbar\omega$, while under the same linear-loss assumption the mean absorption power is
$r\,\langle E\rangle$.  Steady-state balance gives
\(\langle E\rangle = \hbar\omega\,e^{-\gamma},
\)
which underestimates the Planck mean; the geometric cascades are precisely what supply the missing
multiplicative factor $(1-e^{-\gamma})^{-1}$.}{Without them, modeling energy input as a Poisson process where each supra-threshold kick contributes only one \(\hbar\omega\) would yield \(\langle E \rangle = \hbar\omega \cdot e^{-\gamma}\), which underestimates the true result.} The cascades reflect the physical idea that a single large kick can unlock a temporary period of heightened coupling, during which multiple emissions occur in quick succession. This clustering of energy input events is essential to reaching the correct equilibrium distribution. Thus, Planck's law emerges not from quantized field assumptions, but from classical thermodynamic reasoning involving threshold activation, stochastic fluctuation statistics, and asymmetry between input and output paths rooted in thermal accessibility. The kinetic picture complements the variational framework developed in the main text, and together they suggest a robust classical origin for the blackbody spectrum.

\section*{Appendix B: Extremizing the Quantum Free Energy Yields the Schrödinger Equation}
Rewrite Eq.~\ref{eq:FQM} in terms of the probability amplitude \(\psi(x)\)
\begin{equation*}
F_{\text{quantum}}[\psi] = \int \left( U(x) |\psi(x)|^2 + \frac{\hbar^2}{2m} |\nabla \psi(x)|^2 \right) dx,
\end{equation*}
and extremize it subject to the normalization constraint \( \int |\psi(x)|^2 dx = 1 \) to obtain the quantum mechanical states. To enforce normalization, we introduce a Lagrange multiplier \( \lambda \) and define the Lagrangian density:
\begin{equation*}
\mathcal{L}_{\text{QM}} = U(x) |\psi(x)|^2 + \frac{\hbar^2}{2m} |\nabla \psi(x)|^2 - \lambda |\psi(x)|^2.
\end{equation*}

Applying the Euler–Lagrange equation
\(
\frac{\partial \mathcal{L}}{\partial \psi^*} - \nabla \cdot \left( \frac{\partial \mathcal{L}}{\partial (\nabla \psi^*)} \right) = 0,
\)
 we obtain
\begin{equation*}
(U(x) - \lambda)\psi(x) - \frac{\hbar^2}{2m} \nabla^2 \psi(x) = 0,
\end{equation*}
which rearranges to the time-independent Schrödinger equation
\begin{equation*}
\left( -\frac{\hbar^2}{2m} \nabla^2 + U(x) \right) \psi(x) = E \psi(x)
\end{equation*}
where we identify \( \lambda = E \) as the energy eigenvalue. Each eigenfunction \( \psi_n(x) \) corresponds to a stationary state of the system, and the global minimum of the free energy corresponds to the ground state.

The free energy functional in Eq.~\ref{eq:FQM} includes only the real part of the kinetic energy through the Fisher information of \( |\psi(x)| \). The full quantum mechanical energy includes an additional contribution from the phase structure of the wavefunction. Let \( \psi(x) = \sqrt{p(x)} e^{i\theta(x)} \). Then the expectation of the momentum-squared operator is:
\begin{equation*}
\langle \hat{P}^2 \rangle = \hbar^2 \left\langle |\nabla \log \psi(x)|^2 \right\rangle = \frac{\hbar^2}{4} J[p] + \hbar^2 \int p(x) \left( \nabla \theta(x) \right)^2 dx,
\end{equation*}
where \( J[p] = \int \frac{(\nabla p)^2}{p} dx \) is the Fisher information. The total average energy is:
\begin{equation*}
E_{\text{QM}} = \langle U \rangle + \frac{\hbar^2}{8m} J[p] + \frac{\hbar^2}{2m} \int p(x) \left( \nabla \theta(x) \right)^2 dx.
\end{equation*}
This shows that the Fisher-based free energy yields a lower bound on the total energy. The bound is tight when the wavefunction has constant phase (i.e., \( \nabla \theta(x) = 0 \)), in which case the additional term vanishes. In our framework, which optimizes over real-valued probability densities \( p(x) \) without explicit phase structure, the Schrödinger equation still emerges, but the full energy is only recovered in phase-symmetric cases.

Once both contributions to the kinetic energy are included, the free energy becomes exactly the expectation value of the quantum Hamiltonian,
\[
F[\psi] = \langle \psi | \hat{H} | \psi \rangle = \int \psi^*(x) \left( -\frac{\hbar^2}{2m} \nabla^2 + U(x) \right) \psi(x) \, dx.
\]
Minimizing this functional over normalized wavefunctions yields the ground state energy — a procedure known as the Rayleigh–Ritz variational principle. More generally, quantum mechanics identifies all stationary points of \( \langle \hat{H} \rangle \), not just the global minimum. These stationary points correspond to eigenfunctions of \( \hat{H} \), and therefore define all bound states of the system. What makes this formulation distinctive is that all terms in the energy — both kinetic and potential — are expressed in position space, with momentum information encoded in the phase gradient of \( \psi(x) \). This decomposition reveals quantum mechanics as a variational theory over spatial probability amplitudes, where localization and flow are balanced geometrically.

This decomposition also demystifies the appearance of Fisher information in quantum variational formulations: it is not an abstract information measure but simply the amplitude contribution to the kinetic energy. For real-valued wavefunctions with constant phase, the Fisher term accounts for all kinetic energy. More generally, as seen in the Madelung formulation, the total kinetic energy divides into two geometric parts: one associated with spatial localization (the Fisher term), and one with coherent flow (the phase-gradient term). Our variational approach thus reveals how the core structure of quantum mechanics emerges from a balance between these two positional effects, with Fisher information playing a natural role as the localization energy in position space.

\section*{Appendix C: Generalized Free Energy Extremization and Gaussian Solution}
\label{app:appendixC}

We start from our free energy functional \(F_\gamma[p]\) in Eq.~\eqref{eq:general_free_energy}.
To streamline the Euler--Lagrange computation, we rewrite the functional in terms of a real
probability amplitude \(\psi(x)\ge 0\) defined by
\begin{equation*}
p(x) = \psi(x)^2,
\qquad
\int \psi(x)^2\,dx = 1.
\end{equation*}
In one dimension, the Fisher information \(J[p]\) becomes
\begin{equation*}
J[p] = \int \frac{(2\psi\,\partial_x\psi)^2}{\psi^2}\,dx
     = 4\int (\partial_x\psi)^2\,dx,
\end{equation*}
so that
\begin{equation*}
\frac{\hbar^2}{8m}J[p] = \frac{\hbar^2}{2m}\int (\partial_x\psi)^2\,dx.
\end{equation*}
Also, since \(-H[p]=\int p\log p\,dx\), the entropy contribution becomes
\begin{equation*}
-\,k_B T f(\gamma) H[p] = k_B T f(\gamma)\int \psi(x)^2\log\!\bigl(\psi(x)^2\bigr)\,dx.
\end{equation*}
Therefore, the functional can be written as
\begin{equation}
F_\gamma[\psi] =
\int\!\left[
U(x)\psi(x)^2
+ k_B T f(\gamma)\,\psi(x)^2\log\!\bigl(\psi(x)^2\bigr)
+ g(\gamma)\,\frac{\hbar^2}{2m}\,(\partial_x\psi(x))^2
\right]dx.
\label{eq:Fpsi_app}
\end{equation}
To keep \(p(x)\) normalized, we impose normalization with a Lagrange multiplier \(\lambda\):
\begin{equation*}
\widetilde{F}_\gamma[\psi]
=
F_\gamma[\psi] - \lambda\left(\int \psi(x)^2\,dx - 1\right).
\end{equation*}
Next, we apply the Euler--Lagrange (EL) equation
\(
\frac{\partial\mathcal{L}}{\partial\psi} - \partial_x\!\left(\frac{\partial\mathcal{L}}{\partial(\partial_x\psi)}\right)=0
\)
to the integrand of $\widetilde{F}_\gamma[\psi],$ which ignoring the constant \(\lambda\) is:
\begin{equation*}
\mathcal{L}(\psi,\partial_x\psi) =
U\psi^2
+ k_B T f(\gamma)\,\psi^2\log(\psi^2)
+ g(\gamma)\,\frac{\hbar^2}{2m}\,(\partial_x\psi)^2
- \lambda \psi^2.
\label{eq:L_app}
\end{equation*}
Since
\begin{equation*}
\frac{\partial}{\partial\psi}\bigl(\psi^2\log(\psi^2)\bigr)
= 2\psi\bigl(1 + \log(\psi^2)\bigr)
= 2\psi\bigl(1 + 2\log\psi\bigr),
\end{equation*}
the first derivative in the EL equation is
\begin{equation*}
\frac{\partial\mathcal{L}}{\partial\psi}
=
2U\psi
+ 2k_B T f(\gamma)\,\psi\bigl(1 + 2\log\psi\bigr)
- 2\lambda\psi.
\end{equation*}
Similarly,
\begin{equation*}
\frac{\partial\mathcal{L}}{\partial(\partial_x\psi)}
=
g(\gamma)\,\frac{\hbar^2}{m}\,\partial_x\psi,
\qquad
\partial_x\!\left(\frac{\partial\mathcal{L}}{\partial(\partial_x\psi)}\right)
=
g(\gamma)\,\frac{\hbar^2}{m}\,\partial_x^2\psi.
\end{equation*}
Substituting and dividing by \(2\) yields the generalized stationary equation:
\begin{equation}
-\,g(\gamma)\,\frac{\hbar^2}{2m}\,\partial_x^2\psi(x)
+
\left[
U(x)
+ k_B T f(\gamma)\,\bigl(1 + 2\log\psi(x)\bigr)
\right]\psi(x)
= \lambda\,\psi(x).
\label{eq:genEL_app}
\end{equation}

Equation~\eqref{eq:genEL_app} plays the role of a generalized stationary Schr\"odinger equation.
Here \(\lambda\) is the Lagrange multiplier enforcing normalization, and therefore should be
interpreted as a \emph{free-energy eigenvalue} rather than as the physical oscillator energy.
This distinction matters because the entropy term introduces a nonlinear dependence on \(\psi\),
so \(\lambda\) does not coincide with \(\langle E\rangle\) even in the harmonic case. In our
framework, physically meaningful energies are instead computed directly from expectation values
under the extremal distribution (as in the main text), rather than inferred from \(\lambda\).

To compute explicit results for the harmonic oscillator, we now assume the extremal solution
\(p(x)=\psi^2(x)\) is Gaussian with zero mean and variance \(\sigma^2\).  Using the standard
Gaussian values for \(H[p]\) and \(J[p]\), and substituting into Eq.~\eqref{eq:general_free_energy},
the free energy reduces to
\begin{equation}
F(\sigma^2)
=
\frac12 m\omega^2\sigma^2
- \frac12 k_B T f(\gamma)\,\log(2\pi e\,\sigma^2)
+ g(\gamma)\,\frac{\hbar^2}{8m}\,\frac{1}{\sigma^2}.
\label{eq:F_sigma_app}
\end{equation}
To extremize this functional, we differentiate Eq.~\eqref{eq:F_sigma_app} with respect to
\(\sigma^2\), set the result to zero, and obtain a quadratic in \(y=\sigma^2\), whose positive
root yields
\begin{equation}
\sigma^2
=
\frac{k_B T f(\gamma)}{2m\omega^2}
\left[
1 + \sqrt{1 + \frac{g(\gamma)\hbar^2\omega^2}{k_B^2T^2 f(\gamma)^2}}
\right].
\label{eq:sigma_general_app}
\end{equation}
We select the \(+\) branch because it reproduces the classical high-temperature scaling
\(\sigma^2\sim k_B T/(m\omega^2)\).

Next, we specialize Eq.~\eqref{eq:sigma_general_app} to \(f(\gamma)=\gamma\) and
\(\gamma=\hbar\omega/(k_B T)\), which implies
\begin{equation*}
\sigma^2
=
\frac{\hbar}{2m\omega}
\left[
1 + \sqrt{1 + g(\gamma)}
\right].
\end{equation*}
Using Eq.~\eqref{eq:g} (so that \(1+g(\gamma)=\bigl(\tfrac{2}{e^\gamma-1}\bigr)^2\)) gives
Eq.~\eqref{eq:optimal_var}. The mean potential energy in Eq.~\eqref{eq:mean_potential} follows by
direct substitution of \(\sigma^2\) from Eq.~\eqref{eq:optimal_var} into
\(\langle U\rangle=\frac12 m\omega^2\sigma^2\). Finally, invoking the symmetric phase-space
extension described in the main text (so that \(\langle T\rangle=\langle U\rangle\)), the total
mean oscillator energy (Eq.~\eqref{eq:mean_E}) is
\(\langle E\rangle = \frac{\hbar\omega}{2} + \frac{\hbar\omega}{e^\gamma-1}\), and subtracting the
\(T\to 0\) limit isolates the radiative portion
\(\langle E_{\mathrm{rad}}\rangle = \frac{\hbar\omega}{e^\gamma-1}\), which is exactly the Planck
mean energy per mode used in Eqs.~\eqref{eq:PlanckSpectra}--\eqref{eq:PlanckSpectra2}.

\paragraph{Consistency check and the eigenvalue \(\lambda\).}
For completeness, we confirm that the Gaussian ansatz is consistent with the generalized stationary equation
Eq.~\eqref{eq:genEL_app}. Substituting
\(
\psi(x) = (2\pi\sigma^2)^{-1/4}\exp\!\left(-\frac{x^2}{4\sigma^2}\right)
\)
and \(U(x)=\frac12 m\omega^2x^2\) into the left-hand side of Eq.~\eqref{eq:genEL_app} gives
\begin{align*}
\biggl[\, &
g(\gamma)\,\frac{\hbar^2}{2m}\left(\frac{1}{2\sigma^2} - \frac{x^2}{4\sigma^4}\right)
+ \frac{1}{2}m\omega^2 x^2 \\
&\quad + k_B T f(\gamma)\left(1 - \frac{x^2}{2\sigma^2} - \frac{1}{2}\log(2\pi\sigma^2)\right)
\biggr]\psi(x)
= \lambda\,\psi(x).
\end{align*}
Collecting the \(x^2\) terms inside the brackets yields the coefficient
\begin{equation*}
B
=
\frac12 m\omega^2
-\frac12 k_B T f(\gamma)\frac{1}{\sigma^2}
- g(\gamma)\frac{\hbar^2}{8m}\frac{1}{\sigma^4}.
\end{equation*}
Since \(\lambda\) is \(x\)-independent, Eq.~\eqref{eq:genEL_app} holds for all \(x\) if and only if \(B=0\).
But differentiating Eq.~\eqref{eq:F_sigma_app} with respect to \(\sigma^2\) gives
\(\frac{dF}{d(\sigma^2)} = B\), so this EL consistency condition is exactly the same as the extremality
condition used to obtain Eq.~\eqref{eq:sigma_general_app}. With \(B=0\), the remaining constant term fixes
the Lagrange multiplier:
\begin{equation*}
\lambda
=
g(\gamma)\frac{\hbar^2}{4m\sigma^2}
+ f(\gamma)k_B T\left(1 - \frac{1}{2} \log\left(2 \pi \sigma^2\right)\right).
\end{equation*}
Therefore, the extremal Gaussian not only minimizes \(F(\sigma^2)\) but also satisfies the associated
Euler--Lagrange equation, establishing the internal consistency of the variational solution.

\section*{Appendix D: Time-tag realization of the cascade model and analytic counting statistics}
\label{app:simulation}

Appendix~A derives the Planck mean energy per mode from a threshold-activated \emph{emission-cascade} picture:
thermal ``kicks'' arrive at an average rate and only sufficiently energetic kicks can inject radiative quanta.
This appendix is \emph{not} needed for the variational derivation in the main text.  It is included to give one
explicit realization of the Appendix~A mechanism (as requested by the reviewer) and to connect it
to standard photon-counting diagnostics (e.g.\ Hanbury Brown--Twiss bunching) used in single-mode measurements, e.g., in high-$Q$ cavities.

A high-$Q$ cavity provides a single, well-isolated electromagnetic mode with (i) a known resonance frequency
$\omega$, (ii) a linewidth $\kappa$ that sets a natural correlation time $\sim \kappa^{-1}$, and (iii) strong
suppression of multi-mode averaging.  In the microwave single-/few-mode regime, experiments directly measure
thermal (Bose--Einstein) photon statistics and HBT bunching~\cite{HanburyBrownTwiss1956}, as well as the transition
to Poissonian statistics under coherent drive (e.g.\ Refs.~\cite{Goetz2017,Chen2011,Pankratov2025,Keranen2025}).
Our goal here is not to model a specific detector or cavity in detail, but to show that the Appendix~A
\emph{cascade} mechanism admits a minimal stationary time-domain realization with closed-form counting statistics.

Appendix~A assumes kick energies are Boltzmann-distributed and that supra-threshold activation produces
geometrically distributed cascade sizes.  A compact way to reproduce the same burst-size law is to map each kick
energy $E$ into an integer number of quanta,
\[
N=\left\lfloor \frac{E}{\hbar\omega}\right\rfloor,
\qquad
E\sim \mathrm{Exp}(\text{mean }k_B T).
\]
This implies the geometric law on $\{0,1,2,\dots\}$ and its Planck mean:
\begin{align}
\Pr(N\ge n) &= \Pr(E\ge n\hbar\omega)=e^{-n\gamma}, \nonumber \\
\Pr(N=n) &= (1-e^{-\gamma})\,e^{-n\gamma}\qquad(n=0,1,2,\dots), \label{eq:geom0}
\end{align}
so \( \langle N \rangle = \frac{1}{e^\gamma-1} \) where $\gamma=\hbar\omega/(k_B T)$.  Conditioning on $N\ge 1$ recovers the same geometric cascade-size
distribution on $\{1,2,\dots\}$ used in Appendix~A.  Thus Appendix~D and Appendix~A agree on the key steady-state
ingredient: the mean injected energy per kick is $\hbar\omega\,\langle N\rangle=\hbar\omega/(e^\gamma-1)$.

\paragraph{Minimal temporal model (Poisson cluster process).}
Appendix~A does not specify the fine-grained emission times within a cascade; it uses steady-state energy balance
to obtain the Planck mean.  To make time-domain predictions, we choose a minimal model that (i) preserves the
geometric burst sizes in Eq.~\eqref{eq:geom0}, (ii) introduces a single correlation time set by $\kappa^{-1}$, and
(iii) allows cascades to \emph{overlap} in time (a new kick can occur before earlier emissions have all occurred).

We model kick times $\{S_i\}$ as a Poisson process of rate $r$.  Each kick produces a burst size $N_i$ drawn from
Eq.~\eqref{eq:geom0}.  Conditional on $N_i$, we draw independent emission delays
\[
D_{ij}\sim \mathrm{Exp}(\kappa), \qquad j=1,\dots,N_i,
\]
and define photon \emph{time tags} as
\(t_{ij} = S_i + D_{ij}.\)
Thus the photon stream is a standard \emph{Poisson cluster process}.  For simplicity we assume unit detection
efficiency; independent thinning can be added later without changing the normalized expressions below. The $D_{ij}$ are i.i.d.\ \emph{absolute} delays from the parent kick time $S_i$. The \emph{ordered} emission events within cascade $i$ are therefore obtained
by sorting the set $\{t_{ij}\}_{j=1}^{N_i}$, i.e.\ they are the order statistics of i.i.d.\ $\mathrm{Exp}(\kappa)$
samples.  Equivalently, if one prefers a sequential “leakage” construction, the inter-emission waiting times are
independent exponentials with rates $\kappa N_i,\ \kappa(N_i-1),\ \dots,\ \kappa$ (pure-death leakage from an
initial occupancy $N_i$).

Given the time tags $\{t_{ij}\}$ for all photons, for a bin width $\Delta t$ define binned counts
$n_\ell=\#\{t_{ij}\in [\ell\Delta t,(\ell+1)\Delta t)\}$.
We estimate the (binned) second-order coherence for $k\ge 1$ by
\begin{equation*}
g^{(2)}(k\Delta t)=\frac{\langle n_j\,n_{j+k}\rangle}{\langle n_j\rangle^2},
\end{equation*}
and at zero delay by the factorial-moment form
\begin{equation*}
g^{(2)}_{\Delta t}(0)=\frac{\langle n_j(n_j-1)\rangle}{\langle n_j\rangle^2}
=1+\frac{\mathrm{Var}(n_j)-\langle n_j\rangle}{\langle n_j\rangle^2}.
\label{eq:g2b0_def}
\end{equation*}
We also report the Fano factor $F(\Delta t)=\mathrm{Var}(n_j)/\langle n_j\rangle$.
For a Poisson process, $g^{(2)}(\tau)=1$ and $F(\Delta t)=1$ for all $\Delta t$ (with $\tau=k\Delta t$).

\paragraph{Analytic expectation for the cluster realization.}
It is convenient to represent the photon stream as a random ``intensity'' (point process)
\[
I(t)\equiv \sum_{i}\sum_{j=1}^{N_i}\delta\!\left(t-(S_i+D_{ij})\right),
\qquad
\nu \equiv \langle I(t)\rangle = r\,\langle N\rangle ,
\]
where $\langle\cdot\rangle$ denotes an average over realizations. The mean photon flux (intensity) is then
\(
\nu  = \frac{r}{e^\gamma-1}.\) For comparison we also simulate a Poisson photon stream with the same mean rate $\nu$. For $\tau\neq 0$ (so that we exclude the
trivial ``same photon'' contribution at zero delay), the unnormalized two-photon coincidence density is
$\langle I(t)\,I(t+\tau)\rangle$.
Expanding the product gives a sum over all \emph{ordered} photon pairs:
\[
\langle I(t)\,I(t+\tau)\rangle
=
\Biggl\langle
\sum_{i,j}\sum_{i',j'}
\delta\!\left(t-(S_i+D_{ij})\right)\,
\delta\!\left(t+\tau-(S_{i'}+D_{i'j'})\right)
\Biggr\rangle.
\]
There are two conceptually distinct contributions:

\begin{itemize}
\item \textbf{Different kicks ($i\neq i'$):} photons belong to two different cascades.  Since the parent kick
times form a Poisson process and different cascades are independent, these pairs contribute only the
\emph{uncorrelated baseline}
\[
\langle I(t)\rangle\,\langle I(t+\tau)\rangle=\nu^2.
\]

\item \textbf{Same kick ($i=i'$ and $j\neq j'$):} the two photons share the same parent kick time $S_i$, hence
their detection times are correlated through the common shift by $S_i$.  Conditional on a kick at time $S_i=s$,
two independent delays $D_1,D_2$ with density $f_D$ have joint density $f_D(u)f_D(v)$, so the probability
density that their detection-time difference equals $\tau$ is
\[
h(\tau)\equiv \int_{-\infty}^{\infty} f_D(u)\,f_D(u+\tau)\,du,
\]
i.e.\ $h(\tau)$ is the density of the difference $D_1-D_2$ for $D_1,D_2\sim f_D$ i.i.d.
Each kick produces $N_i(N_i-1)$ ordered distinct photon pairs, so averaging over $N$ and over kick times yields
the excess-coincidence term $r\,\langle N(N-1)\rangle\,h(\tau)$.
\end{itemize}

Putting these together gives the explicit decomposition
\[
\langle I(t)\,I(t+\tau)\rangle
=
\nu^2 + r\,\langle N(N-1)\rangle\,h(\tau),
\]
and therefore the normalized second-order coherence is
\begin{equation}
g^{(2)}(\tau)
=
\frac{\langle I(t)\,I(t+\tau)\rangle}{\nu^2}
=
1+\frac{\langle N(N-1)\rangle}{r\,\langle N\rangle^2}\,h(\tau).
\label{eq:g2_general_cluster}
\end{equation}

For our minimal kernel $D\sim\mathrm{Exp}(\kappa)$, $f_D(u)=\kappa e^{-\kappa u}\mathbf{1}_{u\ge 0}$.
A direct integral shows that the difference density is Laplace:
\[
h(\tau)=\int_{-\infty}^{\infty} f_D(u)\,f_D(u+\tau)\,du
=
\frac{\kappa}{2}e^{-\kappa|\tau|}.
\]
For the geometric burst-size law in Eq.~\eqref{eq:geom0}, one also has
$\langle N(N-1)\rangle/\langle N\rangle^2=2$,
so Eq.~\eqref{eq:g2_general_cluster} reduces to the closed form
\begin{equation}
g^{(2)}(\tau)=1+\frac{\kappa}{r}\,e^{-\kappa|\tau|}.
\label{eq:g2_tau_cluster}
\end{equation}
A common ``thermal calibration'' in this minimal model is $r=\kappa$, which yields $g^{(2)}(0)=2$ and reproduces
the standard single-mode thermal (HBT) form $g^{(2)}(\tau)=1+e^{-\kappa|\tau|}$.

For binned counts, $g^{(2)}_{\Delta t}(0)$ is the average of $g^{(2)}(\tau)$ over two times chosen uniformly
within the same bin (equivalently, over the difference of two uniform times in $[0,\Delta t]$):
\[
g^{(2)}_{\Delta t}(0)
=
\frac{1}{(\Delta t)^2}\int_{0}^{\Delta t}\!\!\int_{0}^{\Delta t}
g^{(2)}(u-v)\,du\,dv.
\]
Integrating Eq.~\eqref{eq:g2_tau_cluster} yields
\begin{equation}
g^{(2)}_{\Delta t}(0)
=
1+\frac{2}{r}\,
\frac{\Delta t-\bigl(1-e^{-\kappa\Delta t}\bigr)/\kappa}{(\Delta t)^2}.
\label{eq:g2b0_cluster}
\end{equation}
Finally, using $n^2=n(n-1)+n$ gives
$\mathrm{Var}(n_j)=\langle n_j\rangle+\langle n_j\rangle^2\bigl(g^{(2)}_{\Delta t}(0)-1\bigr)$, hence
\[
F(\Delta t)=\frac{\mathrm{Var}(n_j)}{\langle n_j\rangle}
=
1+\langle n_j\rangle\bigl(g^{(2)}_{\Delta t}(0)-1\bigr).
\]
With $\langle n_j\rangle=\nu\Delta t$ and Eq.~\eqref{eq:g2b0_cluster}, this gives the closed form
\begin{equation}
F(\Delta t)
=
1+\nu\Delta t\left(g^{(2)}_{\Delta t}(0)-1\right)
=
1+\frac{2\nu}{r}\left[1-\frac{1-e^{-\kappa\Delta t}}{\kappa\Delta t}\right],
\qquad
\nu=\frac{r}{e^\gamma-1}.
\label{eq:fano_cluster}
\end{equation}

\begin{figure}[t]
  \centering

  \begin{minipage}{0.8\linewidth}
    \centering
    \textbf{(a)}\\[1mm]
    \includegraphics[width=\linewidth]{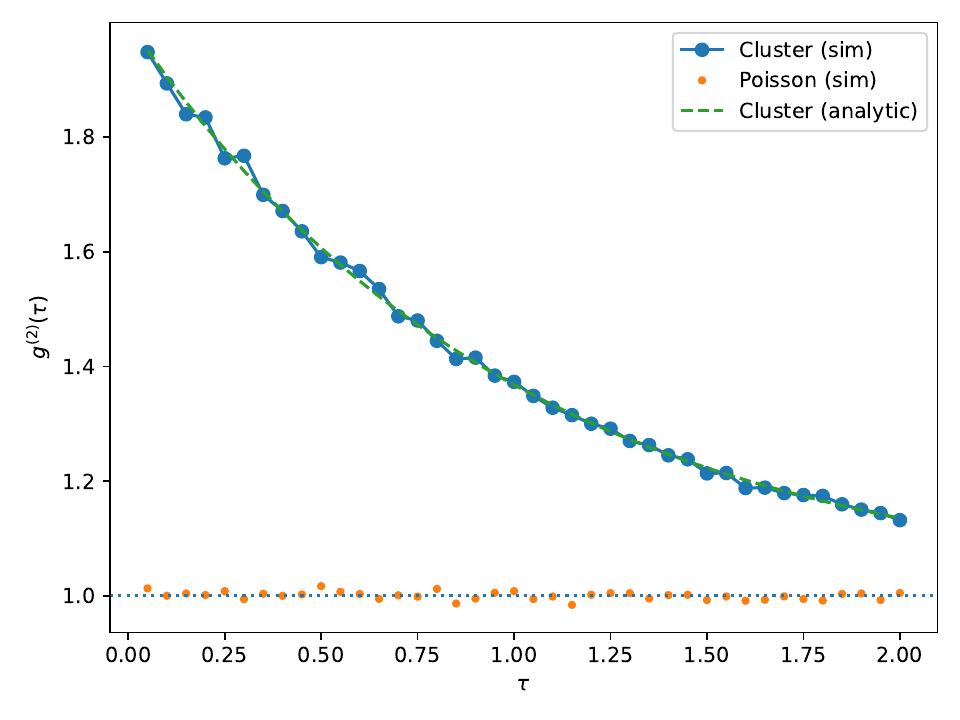}
  \end{minipage}

  \vspace{2mm}

  \begin{minipage}{0.5\linewidth}
    \centering
    \textbf{(b)}\\[1mm]
    \includegraphics[width=\linewidth]{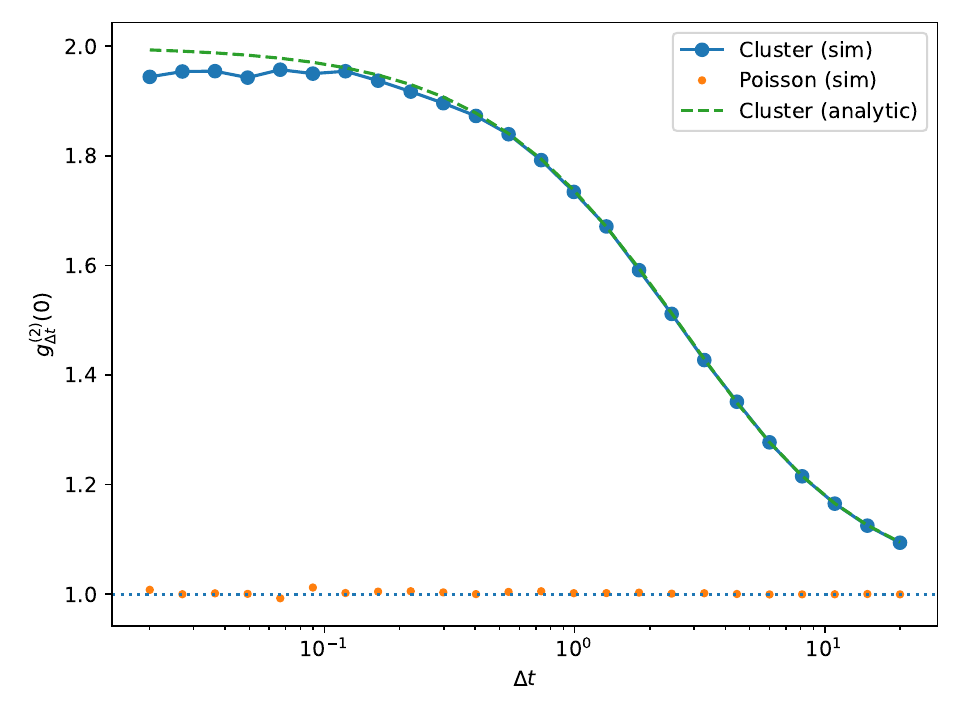}
  \end{minipage}\hfill
  \begin{minipage}{0.5\linewidth}
    \centering
    \textbf{(c)}\\[1mm]
    \includegraphics[width=\linewidth]{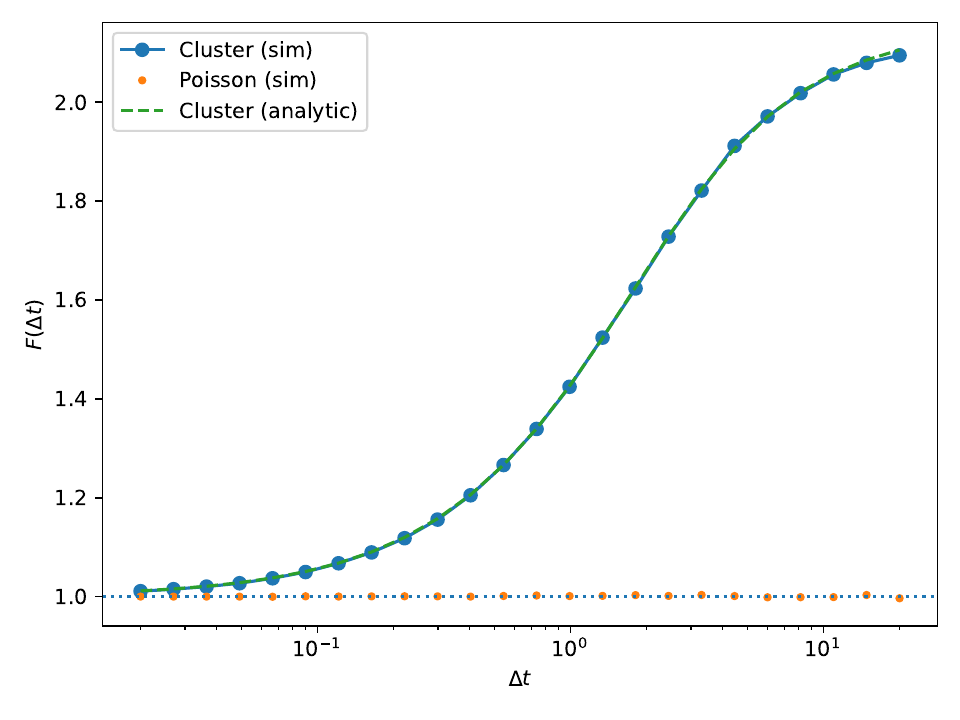}
  \end{minipage}

  \caption{
  Time-domain diagnostics for the bursty cluster realization compared to a matched-mean Poisson
  baseline. Dashed curves show analytic expectations (for the cluster model via Eqs.~\eqref{eq:g2_tau_cluster}--\eqref{eq:fano_cluster}), and markers the simulation results.
  (a) Binned estimate of $g^{(2)}(\tau)$ versus delay $\tau$.
  (b) Binned estimate of $g^{(2)}_{\Delta t}(0)$ versus bin width $\Delta t$.
  (c) Corresponding Fano factor $F(\Delta t)$.
  For Poisson, $g^{(2)}(\tau)=1$ and $F(\Delta t)=1$ for all $\Delta t$.
  }
  \label{fig:appD_panel}
\end{figure}

Figure~\ref{fig:appD_panel} summarizes these diagnostics for the cluster realization (simulation vs analytic)
and compares them to a matched-mean Poisson baseline. Code to reproduce the simulations and this figure is available in a public repository~\cite{GomezUribePlanckFreeEnergyCode}.

\end{document}